\documentclass[prd,twocolumn,showpacs,eqsecnum]{revtex4}

\usepackage{amsmath,amsfonts}
\usepackage{bbm}

\newcommand{\Dov}{D_{\text{ov}}}
\newcommand{\Dw}{D_\text{w}}
\newcommand{\Dt}{\tilde{D}_\text{w}}
\newcommand{\Hw}{H_\text{w}}
\DeclareMathOperator{\sign}{\varepsilon}
\DeclareMathOperator{\tr}{tr}
\DeclareMathOperator{\re}{Re}
\DeclareMathOperator{\diag}{diag}
\DeclareMathOperator{\ind}{index}

\begin{document}

\title{Domain-wall and overlap fermions at nonzero quark chemical
  potential}

\author{Jacques Bloch and Tilo Wettig}
\affiliation{Institute for Theoretical Physics, University of
  Regensburg, 93040 Regensburg, Germany}

\date{September 28, 2007}

\begin{abstract}
  We have recently given a construction of the overlap Dirac operator
  at nonzero quark chemical potential.  Here, we introduce a quark
  chemical potential in the domain-wall fermion formalism and show
  that our earlier result is reproduced if the extent of the fifth
  dimension is taken to infinity and its lattice spacing is taken to
  zero.  We also extend this result to include a bare quark mass,
  consider its continuum limit, and prove a number of properties of
  the overlap operator at nonzero quark chemical potential.  In
  particular, we show that the relation between the anomaly and the
  index of the overlap operator remains valid.
\end{abstract}



\maketitle

\section{Introduction}

The phase diagram of quantum chromodynamics (QCD) as a function of
temperature and chemical potential has been the subject of intense
studies over many years.  It is physically relevant, e.g., for the
study of compact stars, for ultra-relativistic heavy-ion collisions,
and for the physics of the early universe.  The theoretical methods
that have been employed to investigate the QCD phase diagram include
model calculations, effective theories, perturbative studies at high
temperature and density, lattice simulations, and recently also the
AdS/CFT correspondence.  For an overview of the literature and a
summary of the current status, we refer the reader to
Ref.~\cite{Stephanov:2007fk}.

Lattice simulations are the predominant nonperturbative tool to study
QCD from first principles.  The case of nonzero temperature $T$ can be
implemented on the lattice without much effort, and therefore the
temperature dependence of many QCD quantities is very well understood,
see Ref.~\cite{Heller:2006ub} for a review.  This is not true for the
case of nonzero baryon density or, equivalently, quark chemical
potential $\mu$.  The reason is that at $\mu\ne0$, the fermion
determinant becomes complex so that standard importance sampling
methods fail.  This is an example of the so-called sign problem, see
Ref.~\cite{Splittorff:2006vj} for a thorough discussion of this
problem in the context of lattice QCD.  A number of approaches have
been invented to deal with this problem, such as reweighting along the
critical line \cite{Fodor:2001au}, Taylor expansion
\cite{Allton:2002zi}, and analytical continuation from imaginary $\mu$
\cite{deForcrand:2002ci,D'Elia:2002gd}.  Using these approaches,
attempts have been made to determine the transition line $T_c(\mu)$
and to locate the critical end-point, see Ref.~\cite{Schmidt:2006us}
for a review.  While the results are encouraging, the sign problem
remains unsolved in principle.

Most of the lattice results for QCD thermodynamics have been obtained
with staggered fermions, which reduce the fermion doubling problem and
implement a remnant chiral symmetry on the lattice.  For simulations
with less than four staggered flavors, the so-called rooting problem
has been the subject of some debate, see Ref.~\cite{Sharpe:2006re} for
a review.  We have no intention to enter this debate here.  Rather,
our aim is to investigate how a quark chemical potential can be
implemented in a fermion operator that implements an exact chiral
symmetry on the lattice.

In Ref.~\cite{Bloch:2006cd} we showed how this can be done for the
overlap operator \cite{Narayanan:1994gw,Neuberger:1997fp}.  (For
earlier work with a similar focus, see
Refs.~\cite{Bietenholz:1998rj,Bietenholz:1999km}.)  The resulting
operator, $\Dov(\mu)$, contains the sign function of a nonhermitian
matrix.  We also showed that quenched lattice results obtained with
this operator agree with analytical predictions from nonhermitian
chiral random matrix theory at $\mu\ne0$
\cite{Splittorff:2003cu,Osborn:2004rf,Akemann:2004dr}, see also
\cite{Akemann:2003wg}.  Furthermore, the authors of
Ref.~\cite{Gattringer:2007uu} have shown that our construction of
$\Dov(\mu)$ yields the correct energy density for free fermions.

In the present paper, we introduce a quark chemical potential in the
domain-wall fermion formalism
\cite{Callan:1984sa,Kaplan:1992bt,Shamir:1993zy,Furman:1994ky}, which
can be viewed as a particular truncation of the overlap operator.  In
analogy to the well-known result at $\mu=0$, we show that in the limit
in which the extent of the fifth dimension is taken to infinity and
its lattice spacing is taken to zero, our earlier result for
$\Dov(\mu)$ is reproduced.  We also extend this result to include a
bare quark mass, consider its continuum limit, and prove several
properties of $\Dov(\mu)$, including the relation between the anomaly
and the index.

We should remark that at present, the topic we address here may seem
to be mainly of theoretical interest since lattice simulations with
such an operator, especially at $\mu\ne0$, are numerically much more
expensive than those with staggered fermions.  However, as computers
and algorithms improve, more and more lattice QCD simulations will be
done with overlap and domain-wall fermions.  As a first step towards
such simulations at $\mu\ne0$, we have already proposed and tested a
new iterative method to compute the sign function of nonhermitian
matrices \cite{Bloch:2007aw}.

This paper is organized as follows.  In Sec.~\ref{sec:dwf} we review
domain-wall fermions at $\mu=0$, extend the domain-wall action to
$\mu\ne0$, and consider the limit of this action for infinite extent
and zero lattice spacing of the fifth dimension.  After a short side
remark in Sec.~\ref{sec:remark}, we discuss the continuum limit of
$\Dov(\mu)$ in Sec.~\ref{sec:cont}.  In Sec.~\ref{sec:prop} we prove a
number of properties of $\Dov(\mu)$ that were stated, but not proven,
in Ref.~\cite{Bloch:2006cd}.  We conclude with a summary and outlook
in Sec.~\ref{sec:summary}.

\section{Domain-wall fermions at $\mu\ne0$ and relation to
  overlap operator}
\label{sec:dwf}

\subsection{Domain-wall fermions at \boldmath{$\mu=0$}}

We start with the definition of the Wilson Dirac operator in four
dimensions, which is given by \cite{Hasenfratz:1983ba}
\begin{align}
  \label{eq:dwmu}
  \Dw(\mu)&=(4+M)-\frac12\sum_{i=1}^3\left(T_i^++T_i^-\right)\notag\\
  &\quad-\frac12\left(e^\mu T_4^++e^{-\mu} T_4^-\right)
  \intertext{with}
  (T_\nu^\pm)_{xy}&=(1\pm\gamma_\nu)U_{\pm\nu}(x)\delta_{y,x\pm\hat\nu}\:,
\end{align}
where $M$ is the Wilson mass, the $U\in\text{SU(3)}$ are the lattice
gauge fields with $U_{-\nu}(x)=U_\nu^\dagger(x-\hat\nu)$, the
$\gamma_\nu$ are the usual Euclidean Dirac matrices, the Wilson
parameter $r$ has been fixed at $r=1$, the 4-d lattice spacing $a$ has
been set to unity, and for later convenience we have already included
a quark chemical potential $\mu$.

Domain-wall fermions
\cite{Callan:1984sa,Kaplan:1992bt,Shamir:1993zy,Furman:1994ky} are
constructed by introducing an additional fifth dimension with lattice
spacing $a_5$ and extent $L_s$.  The fermion fields now have an
additional index $s=1,\ldots,L_s$, while the gauge fields remain
four-dimensional and do not depend on $s$.  At $\mu=0$, the
domain-wall fermion action is given by
\cite{Shamir:1993zy,Furman:1994ky}
\begin{align}
  \label{eq:dw}
  -S_5&=\bar\psi D_5\psi\notag\\
  &=\sum_{s=1}^{L_s}(\bar\psi_sA\psi_s-\bar\psi_sP_R\psi_{s+1}
  -\bar\psi_sP_L\psi_{s-1})\:,
\end{align}
where $A=a_5\Dw(\mu=0)+1$, the chiral projection operators $P_R$ and
$P_L$ are defined by $P_{R/L}=\frac12(1\pm\gamma_5)$, and the fermion
fields satisfy the following boundary conditions in the fifth
dimension,
\begin{align}
  \label{eq:bc}
  P_R\psi_{L_s+1}=-mP_R\psi_1\:,\quad
  P_L\psi_0=-mP_L\psi_{L_s}\:.
\end{align}
The quantity $m$ is a bare quark mass parameter.  In
Eq.~\eqref{eq:dw}, the Wilson mass has to be in the range $-2<M<0$ to
obtain renormalizable solutions in the fifth dimension and to avoid
the existence of doublers \cite{Kaplan:1992bt,Shamir:1993zy}.

To take the $L_s\to\infty$ limit, the domain-wall fermion action of
Eq.~\eqref{eq:dw} is supplemented by a pseudo-fermion action to cancel
divergences due to the heavy fermions in the large-$L_s$ limit
\cite{Narayanan:1992wx,Frolov:1992ck,Frolov:1993zr,Narayanan:1993sk,Furman:1994ky,Vranas:1997da,Neuberger:1997bg}.
We use the pseudo-fermion action of Ref.~\cite{Vranas:1997da}, which
is given by Eq.~\eqref{eq:dw} with anti-periodic boundary conditions,
i.e., $m=1$ in Eq.~\eqref{eq:bc}, but in which bosonic fields are used
instead of the fermionic ones.  As the fermion action is only defined
up to a constant normalization factor, we choose to multiply the
pseudo-fermion action by $1/2$.

\subsection{Domain-wall fermions at \boldmath{$\mu\ne0$}}

We define the domain-wall fermion action at $\mu\ne0$ to be the same
action as in Eq.~\eqref{eq:dw}, except that $\Dw(0)$ is replaced by
$\Dw(\mu)$.  With very minor modifications, the arguments of
Ref.~\cite{Kaplan:1992bt} leading to the bounds on the Wilson mass
apply to the case of $\mu\ne0$ as well, and we again obtain the
requirement $-2<M<0$.

\subsection{\boldmath{$L_s\to\infty$} limit of domain-wall
  fermions at \boldmath{$\mu\ne0$}}

For $\mu=0$, the connection between domain-wall fer\-mions and the
overlap operator in the $L_s\to\infty$ limit has been exhibited in a
number of earlier works, e.g.,
Refs.~\cite{Neuberger:1997bg,Kikukawa:1999sy,Borici:1999zw,Edwards:2000qv}.
To make the presentation self-contained, we now retrace some of the
steps taken in these papers, in particular
Refs.~\cite{Borici:1999zw,Edwards:2000qv}, with small modifications
suitable for our purposes.

The idea is to introduce successive spinor transformations to
diagonalize the Dirac operator in the fifth dimension and integrate
out the fermion fields.  To this end, we start with the transformation
\begin{equation}
  \label{eq:trans1}
    \psi_s = \left\{
      \begin{array}{ll}
         P_R \chi_s + P_L \chi_{s+1} & \text{for } 1\le s\le L_s-1\:,\\
         P_R \chi_{L_s} + P_L \chi_1 & \text{for } s=L_s\:.
      \end{array}\right.
\end{equation}
It is straightforward to show that this transformation is orthogonal,
with Jacobian equal to 1.  Substituting Eq.~\eqref{eq:trans1} into
Eq.~\eqref{eq:dw} with boundary conditions \eqref{eq:bc} yields
\begin{align}
  \label{eq:dw2}
  -S_5&=\sum_{s=2}^{L_s}\bar\psi_s(AP_R-P_L)\chi_s
  +\bar\psi_1(AP_R+mP_L)\chi_1\notag\\
  &+\sum_{s=2}^{L_s}\bar\psi_{s-1}(AP_L-P_R)\chi_s+
  \bar\psi_{L_s}(AP_L+mP_R)\chi_1\:.
\end{align}
To simplify the first term in Eq.~\eqref{eq:dw2}, we introduce the
transformation
\begin{equation}
  \label{eq:trans2}
  \bar\psi_s=\bar\chi_s(AP_R-P_L)^{-1}\:,
\end{equation}
which is diagonal in the fifth dimension.  The nontrivial Jacobian of
this transformation can be ignored since it is cancelled by the
corresponding Jacobian for the pseudo-fermions.  We also define an
operator $T$ by
\begin{equation}
  \label{eq:t1}
  T=-(AP_R-P_L)^{-1}(AP_L-P_R)\:,
\end{equation}
which is the transfer matrix in the fifth dimension
\cite{Neuberger:1997bg} and will be discussed in more detail below.
Using Eqs.~\eqref{eq:trans2} and \eqref{eq:t1}, Eq.~\eqref{eq:dw2}
becomes
\begin{align}
  \label{eq:dw3}
  -S_5&=\sum_{s=2}^{L_s}\bar\chi_s\chi_s
  +\bar\chi_1(P_R-mP_L)\chi_1\notag\\
  &-\sum_{s=2}^{L_s}\bar\chi_{s-1}T\chi_s
  -\bar\chi_{L_s}T(P_L-mP_R)\chi_1\:,
\end{align}
where we have used 
\begin{align}
  (AP_R-P_L)^{-1}AP_R&=P_R\:,&
  (AP_R-P_L)^{-1}P_L&=-P_L\:,\notag\\
  (AP_L-P_R)^{-1}AP_L&=P_L\:,&
  (AP_L-P_R)^{-1}P_R&=-P_R\:.
\end{align}
(See below for comments on the invertibility of $AP_R-P_L$ and
$AP_L-P_R$.)  The structure of Eq.~\eqref{eq:dw3} suggests to
transform the $\bar\chi_s$ for $s>1$ according to
\begin{align}
  \label{eq:trans3}
  \bar\eta_{s}&=\bar\chi_{s}-\bar\chi_{s-1}T\:,
  \intertext{with inverse transformation}
  \label{eq:trans3inv}
  \bar\chi_s&=\bar\chi_1T^{s-1}+\sum_{i=2}^s\bar\eta_iT^{s-i}\:.
  \intertext{The $\chi_s$ for $s>1$ are transformed according to}
  \label{eq:trans4}
  \chi_{s}&=\eta_{s}+T^{L_s+1-s}(P_L-mP_R)\chi_1\:.
\end{align}
Both of these transformations have a Jacobian equal to 1.
Inserting Eqs.~\eqref{eq:trans3}, \eqref{eq:trans3inv}, and
\eqref{eq:trans4} into Eq.~\eqref{eq:dw3} leads to
\begin{align}
  -S_5&=\sum_{s=2}^{L_s}\bar\eta_s\eta_s
  +\bar\chi_1D_4\chi_1
  \intertext{with}
  D_4&=P_R-mP_L-T^{L_s}(P_L-mP_R)\:.
\end{align}
The $\eta_s$ and $\bar\eta_s$ can now be integrated out trivially.
Finally, we integrate out $\chi_1$ and $\bar\chi_1$ and obtain,
together with the corresponding contribution of the pseudo-fer\-mions,
\begin{align}
  \frac{\det D_4(m)}{\det \frac12D_4(1)}=\det D_\text{eff}
\end{align}
with an effective 4-d operator $D_\text{eff}$ given by
\begin{align}
  \label{eq:deff}
  D_\text{eff}=(1+m)+(1-m)\gamma^5\frac{1-T^{L_s}}{1+T^{L_s}}\:.
\end{align}

We now take a closer look at the transfer matrix of Eq.~\eqref{eq:t1},
which can be rewritten as
\begin{align}
  \label{eq:t2}
  T=(1+a_5\Hw P_R)^{-1}(1-a_5\Hw P_L)
\end{align}
with $\Hw=\gamma_5\Dw$.  For $\mu=0$, $\Dw$ is $\gamma_5$-hermitian,
i.e., it satisfies $\Dw^\dagger=\gamma_5\Dw\gamma_5$, and thus $\Hw$
is hermitian.  From this it follows that $T$ is also hermitian.  The
transfer matrix can be related to a 4-d Hamiltonian $H_t$ by writing
it in the form
\begin{align}
  \label{eq:tht}
  T=\frac{1-a_5H_t}{1+a_5H_t}
\end{align}
with
\begin{align}
  \label{eq:ht}
  H_t=(2+a_5\Hw\gamma_5)^{-1}\Hw=\Hw(2+a_5\gamma_5\Hw)^{-1}\:.
\end{align}
For $\mu=0$, $H_t$ is hermitian. 

Up to this point, everything went through as in
Refs.~\cite{Borici:1999zw,Edwards:2000qv}.  We now move on to the case
of $\mu\ne0$, in which $\Dw$ ceases to be $\gamma_5$-hermitian.
Therefore, neither $\Hw$ nor $H_t$ nor $T$ are hermitian.  To obtain
the $L_s\to\infty$ limit of Eq.~\eqref{eq:deff}, we consider the
matrix function
\begin{align}
  f(H_t)=\frac{1-T^{L_s}}{1+T^{L_s}}\:,
\end{align}
where $T$ is given by Eq.~\eqref{eq:tht} with a nonhermitian matrix
$H_t$.  A function $f$ of an arbitrary complex matrix $C$ can be
defined by \cite{Golub1989}
\begin{align}
  \label{eq:specdec}
  f(C)=\frac1{2\pi i}\oint_\Gamma dz f(z)\left[z-C\right]^{-1}\:,
\end{align}
where the integral is defined component-wise and $\Gamma$ is a
collection of closed contours in $\mathbb C$ such that $f$ is analytic
inside and on $\Gamma$ and such that $\Gamma$ encloses the spectrum of
$C$.  We are therefore interested in
\begin{align}
  \sigma=\frac{1-t^{L_s}}{1+t^{L_s}}\:,
\end{align}
where
\begin{align}
  t=\frac{1-z}{1+z}
\end{align}
and $z\in\mathbb C$ is an eigenvalue of $a_5H_t$.  If $|t|<1$
($|t|>1$), $\sigma\to1$ ($\sigma\to-1$) as $L_s\to\infty$, i.e., we
can write $\sigma\to\sign(1-|t|^2)$, where $\sign$ denotes the sign
function.  For $z\in\mathbb R$, $|t|<1$ ($|t|>1$) if $z>0$ ($z<0$),
and hence $\sigma=\sign(z)$.  Let us now consider $z\in\mathbb C$ with
$x=\re z$, for which
\begin{align}
  1-|t|^2=\frac{4x}{|1+z|^2}\:.
\end{align}
From this expression we obtain immediately
\begin{align}
  \label{eq:sign1}
  \lim_{L_s\to\infty}\sigma=\sign(x)=\sign(\re z)=:\sign(z)\:,
\end{align}
where the last equality defines the sign function of a complex number.
This can also be written as
\begin{align}
  \label{eq:sign2}
  \sign(z)=\frac z{\sqrt{z^2}}
\end{align}
with the branch cut of the square root along the negative real axis.
We thus obtain
\begin{align}
  \label{eq:Deff}
  \lim_{L_s\to\infty} D_\text{eff}(\mu)=(1+m)
    +(1-m)\gamma_5\sign(H_t(\mu))\:,
\end{align}
where the sign function $\sign(C)$ of a nonhermitian matrix $C$ is
defined formally by Eq.~\eqref{eq:specdec} in combination with
Eq.~\eqref{eq:sign1} or \eqref{eq:sign2}.  A simpler form for
$\sign(C)$ can be obtained if $C$ can be diagonalized, i.e., if it can
be written in the form $U\Lambda U^{-1}$ with $U\in\text{Gl}(N,\mathbb
C)$ and $\Lambda=\diag(\lambda_1,\ldots,\lambda_N)$, where $N$ is the
dimension of $C$.  It then follows from Eq.~\eqref{eq:specdec} that
\begin{align}
  f(C)&=Uf(\Lambda)U^{-1}
  \intertext{with}
  f(\Lambda)&=\diag(f(\lambda_i))
\end{align}
so that the matrix sign function can be defined by \cite{Roberts1980}
\begin{align}
  \label{eq:msign}
  \sign(C)=U\sign(\re\Lambda)\,U^{-1}\:.
\end{align}
Note that if $C$ cannot be diagonalized, one can use the Jordan
canonical form instead, see Ref.~\cite{Bloch:2007aw} for details.

In the course of the derivation, we have assumed (i) that the
operators $AP_R-P_L$ and $AP_L-P_R$ are invertible and (ii) that the
elements of the diagonal matrix $\Lambda$ are not zero or purely
imaginary so that the sign function is well-defined.  For any of these
assumptions to be violated, the gauge-field would have to be
fine-tuned.  This happens on a gauge field set of measure zero and can
therefore be ignored in practice.  The same remark applies to the
possibility that $H_t(\mu)$ might not be diagonalizable.

\subsection{\boldmath{$a_5\to0$} limit}

To recover the standard overlap operator, it remains to take the limit
$a_5\to0$ in Eq.~\eqref{eq:Deff}, and therefore in Eq.~\eqref{eq:ht},
which yields
\begin{align}
  \Dov(\mu)&=\lim_{a_5\to0}\lim_{L_s\to\infty}D_\text{eff}(\mu)
  \notag\\
  &=(1+m)+(1-m)\gamma_5\sign(\Hw(\mu))\:.
  \label{eq:dov}
\end{align}
For $m=0$, Eq.~\eqref{eq:dov} together with Eq.~\eqref{eq:msign}
agrees with our earlier result \cite{Bloch:2006cd}.  For $m\ne0$, we
see that the quark mass is included in the same way as for $\mu=0$
\cite{Neuberger:1997bg}.

\section{Side remark}
\label{sec:remark}

Note that for $\mu=0$, there is an expression for the overlap operator
that is equivalent to Eq.~\eqref{eq:dov}, i.e.,
\begin{align}
  \label{eq:alt}
  \Dov=(1+m)+(1-m)\frac{\Dw}{\sqrt{\Dw^\dagger\Dw}}\:.
\end{align}
One could be tempted to use this expression for $\mu\ne0$ as well,
with $\Dw$ replaced by $\Dw(\mu)$.  However, the resulting operator is
not equivalent to Eq.~\eqref{eq:dov} for $\mu\ne0$ due to the lack of
$\gamma_5$-hermiticity of $\Dw(\mu)$.  In particular, it does not
satisfy the Ginsparg-Wilson condition and has no exact zero modes at
finite lattice spacing.  Thus, the expression \eqref{eq:alt} is not
suitable for an extension to $\mu\ne0$.

\section{Continuum limit of \boldmath{$\Dov(\mu)$}}
\label{sec:cont}

In this (and only this) section we reintroduce the 4-d lattice spacing
$a$ that was set to unity earlier and write the Wilson Dirac operator,
with the argument $\mu$ suppressed, in the form
\cite{Hernandez:1998et}
\begin{align}
  \Dw=\frac1a\bigl[-(1+s)+a\Dt\bigr]\:,
\end{align}
where $1+s=-Ma$ with $|s|<1$ and $\Dt$ is the massless Wilson
operator, $\Dt=\Dw(M=0)$.  We therefore have for $\Hw=\gamma_5\Dw$
\begin{align}
  (a\Hw)^2&=(1+s)^2 -a(1+s)(\Dt+\gamma_5\Dt\gamma_5)
  +\mathcal O(a^2)\notag\\
  &=(1+s)^2+\mathcal O(a^2)\:,
\end{align}
where the last step follows from the fact that $\Dt$ anticommutes with
$\gamma_5$ up to terms of order $a$.  Using Eqs.~\eqref{eq:specdec}
and \eqref{eq:sign2}, the matrix sign function can be written as
\begin{align}
  \sign(\Hw)=\frac{\Hw}{\sqrt{\Hw^2}}\:,
\end{align}
and we find
\begin{align}
  \gamma_5\sign(\Hw)=-1+\frac a{1+s}\Dt+\mathcal O(a^2)\:.
\end{align}
Equation~\eqref{eq:dov} thus becomes
\begin{align}
  \Dov(\mu)&=\frac1a[(1+ma)+(1-ma)\gamma_5\sign(\Hw)]\notag\\
  &=2m+\frac1{1+s}\Dt(\mu)+\mathcal O(a)\:.
\end{align}
Since for $a\to0$ the Wilson Dirac operator becomes the continuum
Dirac operator, this is also true for $\Dov(\mu)$, up to a
normalization factor.  There was no need in the above derivation to
use the $\gamma_5$-hermiticity of $\Dt$, which is lacking for
$\mu\ne0$.  The only input required was the fact that $\Dt$
anticommutes with $\gamma_5$ in the continuum limit, which holds for
$\mu\ne0$ as well.

\section{Properties of \boldmath{$\Dov(\mu)$}}
\label{sec:prop}

In Ref.~\cite{Bloch:2006cd} we stated, but did not prove, a number of
properties of $\Dov(\mu)$ at $m=0$.  We supply the missing proofs
here, assuming $m=0$ throughout this section.

Property 1: For $\mu\ne0$, $\Dov(\mu)$ is no longer
$\gamma_5$-hermitian but satisfies
\begin{align}
  \label{eq:prop1}
  \gamma_5\Dov(\mu)\gamma_5=\Dov^\dagger(-\mu)\:
\end{align}
instead.  From Eq.~\eqref{eq:dov} with $m=0$ we obtain
\begin{align}
  \gamma_5\Dov(\mu)\gamma_5
  &=1+\sign(\Hw(\mu))\gamma_5\:,\notag\\
  \Dov^\dagger(-\mu)
  &=1+\sign^\dagger(\Hw(-\mu))\gamma_5\:,
\end{align}
from which Eq.~\eqref{eq:prop1} follows because of
\begin{align}
  \sign^\dagger(\Hw(-\mu))&=\sign^\dagger(\gamma_5\Dw(-\mu))
  =\sign^\dagger(\Dw(\mu)^\dagger\gamma_5)\notag\\
  &=\sign(\gamma_5\Dw(\mu))=\sign(\Hw(\mu))\:.
\end{align}
In the second step, we used the well-known fact that $\Dw$ satisfies
Eq.~\eqref{eq:prop1}, which follows from Eq.~\eqref{eq:dwmu} and from
$\gamma_5T_\nu^\pm\gamma_5=(T_\nu^\mp)^\dagger$.  In the third step,
we used the fact that the sign function satisfies
$\sign(C^\dagger)=\sign^\dagger(C)$ for any matrix $C$, which follows
from Eq.~\eqref{eq:specdec} with $\sign^*(z)=\sign(z^*)$, or from
Eq.~\eqref{eq:msign}.  Note that the proof of Property 1 also goes
through for $m\ne0$.

Property 2: $\Dov(\mu)$ satisfies a Ginsparg-Wilson relation
\cite{Ginsparg:1981bj} of the form
\begin{align}
  \label{eq:gwr}
  \{D,\gamma_5\}=D\gamma_5D\:.
\end{align}
Setting $m=0$ in Eq.~\eqref{eq:dov} and using the shorthand $\sign$
for $\sign(\Hw(\mu))$, we have
\begin{align}
  \{\Dov(\mu),\gamma_5\}&=2\gamma_5+
  \sign+\gamma_5\sign\gamma_5\:,\notag\\
  \Dov(\mu)\gamma_5\Dov(\mu)&=
  (1+\gamma_5\sign)\gamma_5(1+\gamma_5\sign)\notag\\
  &=\gamma_5+\gamma_5\sign^2+\sign+\gamma_5\sign\gamma_5\:.
\end{align}
Equation~\eqref{eq:gwr} now follows immediately from $\sign^2(C)=1$
for any matrix $C$, which in turn follows from Eq.~\eqref{eq:specdec}
with $\sign^2(z)=1$, or from Eq.~\eqref{eq:msign}.

Property 3: All eigenvalues of $\Dov(\mu)$ that are not equal to 0 or
2 come in pairs $\lambda$ (with eigenvector $\psi$) and
$\lambda/(\lambda-1)$ (with eigenvector $\gamma_5\psi$).  We first
note, with all arguments suppressed, that
\begin{align}
  \Dov\psi&=(1+\gamma_5\sign)\psi=\lambda\psi
  \intertext{implies}
  \sign\psi&=(\lambda-1)\gamma_5\psi\:,
\end{align}
from which we conclude that
\begin{align}
  \Dov(\gamma_5\psi)&=\gamma_5\psi+\gamma_5\sign\gamma_5\psi
  =\gamma_5\psi+\gamma_5\sign\frac{\sign\psi}{\lambda-1}\notag\\
  &=\frac{\lambda}{\lambda-1}(\gamma_5\psi)\:,
\end{align}
where we have again used $\sign^2=1$.  Note that for $\mu=0$, all
eigenvalues lie on the circle with radius 1 and center at 1, in which
case $\lambda$ and $\lambda/(\lambda-1)$ are complex conjugates of
each other.

Property 4: The mapping $\lambda\to z=2\lambda/(2-\lambda)$ projects
the pair $\lambda$ and $\lambda/(\lambda-1)$ to a complex conjugate
pair $\pm z$.  This follows from elementary algebra.

Property 5 makes statements about the eigenvectors of $\Dov(\mu)$
corresponding to eigenvalue 0 or 2.  In these cases, Property 3
implies that $\psi$ and $\gamma_5\psi$ are degenerate eigenvectors of
$\Dov(\mu)$.  This means that $\gamma_5$ commutes with $\Dov(\mu)$ in
the corresponding degenerate subspace and can thus be diagonalized in
this subspace.  Because of $\gamma_5^2=1$ the eigenvalues of
$\gamma_5$ are $\pm1$, i.e., the eigenvectors of $\Dov(\mu)$
corresponding to $\lambda=0$ or 2 can be arranged to have definite
chirality.  In the following we denote by $n_\lambda^\pm$ the number
of eigenvectors corresponding to $\lambda=0$ or 2 with
$\langle\gamma_5\rangle=\pm1$.  Consider now the operator
$B=\Dov+\gamma_5\Dov\gamma_5$, where the argument $\mu$ has been
suppressed.  It is easily shown that if $\psi_\lambda$ is an
eigenvector of $\Dov$ with eigenvalue $\lambda$, then $\psi_\lambda$
and $\gamma_5\psi_\lambda$ are degenerate eigenvectors of $B$ with
eigenvalue $\lambda^2/(\lambda-1)$.  For $\lambda\ne0,2$, we now
construct the vectors
$\psi_\lambda^\pm=\psi_\lambda\pm\gamma_5\psi_\lambda$.  According to
Property 3, $\psi_\lambda$ and $\gamma_5\psi_\lambda$ are linearly
independent in this case, and therefore the two vectors
$\psi_\lambda^\pm$ are nonzero and linearly independent.  Moreover,
they are also eigenvectors of $\gamma_5$ with eigenvalue $\pm1$,
respectively.  We now consider $B$ in a basis consisting of the
$\psi_\lambda^\pm$ and of the eigenvectors of $\Dov$ corresponding to
$\lambda=0$ and 2 with definite chirality.  Since in this basis the
operators $\gamma_5$ and $B$ are simultaneously diagonal with
eigenvalues $\lambda^{(\gamma_5)}_i$ and $\lambda^{(B)}_i$,
respectively, we have
\begin{align}
  \tr&(\gamma_5B)=\sum_i\lambda^{(\gamma_5)}_i\lambda^{(B)}_i\notag\\
  &=\frac12\sum_{\lambda_i\ne0,2}(d_i-d_i)\frac{\lambda_i^2}{\lambda_i-1}
  +(n_0^+-n_0^-)0+(n_2^+-n_2^-)4\notag\\
  &=4(n_2^+-n_2^-)\:,
\end{align}
where in the second line $d_i$ is the (accidental) degeneracy of the
eigenvalue $\lambda_i\ne0,2$ of $\Dov$ and the factor of $1/2$ in
front of the sum removes a double counting of eigenvalues.  An
analogous argument holds for $\tilde D_\text{ov}=2-\Dov$, which also
satisfies the Ginsparg-Wilson relation and for which the roles of
$\lambda=0$ and $\lambda=2$ are interchanged, leading to
\begin{align}
  \tr(\gamma_5\tilde B)=4(n_0^+-n_0^-)\:,
\end{align}
where $\tilde B=\tilde D_\text{ov}+\gamma_5\tilde
D_\text{ov}\gamma_5$.  From $\tr(\gamma_5\tilde B)=-\tr(\gamma_5B)$ we
conclude that
\begin{align}
  n_0^+-n_0^-&=-(n_2^+-n_2^-)
\end{align}
as stated in Ref.~\cite{Bloch:2006cd}.  From
$\tr(\gamma_5B)=2\tr(\gamma_5\Dov)$ we also conclude that
\begin{align}    
  \label{eq:index}
  -\tr(\gamma_5\Dov)=2(n_0^+-n_0^-)=2\ind(\Dov)\:.
\end{align}
The relation \eqref{eq:index} between the anomaly and the index of
$\Dov$ was already proven for $\mu=0$ in
Refs.~\cite{Narayanan:1993sk,Hasenfratz:1998ri,Luscher:1998pqa}.  Our
simple derivation shows that it remains valid at \mbox{$\mu\ne0$}.
(The method introduced in Ref.~\cite{Luscher:1998pqa} also works at
$\mu\ne0$ without modifications.)  Eq.~\eqref{eq:index} was used in
Ref.~\cite{Bloch:2006cd} to explain an observed shift in the number of
zero modes of $\Dov(\mu)$ as a function of $\mu$.

Property 6 concerns the normality of $\Dov(\mu)$.  From
Eqs.~\eqref{eq:prop1} and \eqref{eq:gwr}, one easily shows that
\begin{equation}
  \label{eq:normal}
  \begin{split}
  \Dov(\mu)\Dov^\dagger(-\mu)&=\Dov(\mu)+\Dov^\dagger(-\mu)\\
  &=\Dov^\dagger(-\mu)\Dov(\mu)\:.
  \end{split}
\end{equation}
This means that for $\mu=0$, $\Dov$ is a normal operator, whereas for
$\mu\ne0$, we cannot conclude anything from Eq.~\eqref{eq:normal}
about the normality of $\Dov(\mu)$.  This suggests that for $\mu\ne0$,
$\Dov(\mu)$ is not a normal operator (at least generically).  This
expectation is confirmed numerically.  It is interesting to note that
the operator \eqref{eq:normal} is equal to the operator $B$ we
defined in the proof of Property 5.

\section{Summary}
\label{sec:summary}

We have extended the domain-wall formalism to non\-zero quark chemical
potential and have shown that in the limit in which $L_s\to\infty$ and
$a_5\to0$ we obtain an expression for the overlap Dirac operator that
is identical to our earlier result \cite{Bloch:2006cd}.  We have also
included a bare quark mass, considered the continuum limit, and proven
a number of analytical properties of this operator.

In actual lattice simulations, the use of Eq.~\eqref{eq:dov} will be
hindered by two problems.  The first is the infamous sign problem that
plagues lattice QCD at $\mu\ne0$.  We have nothing new to say about
this problem here.  The second problem is that not much is known about
efficient numerical computations of the sign function of a
nonhermitian matrix.  As remarked earlier, we have started to address
the second problem in Ref.~\cite{Bloch:2007aw}.

Work is in progress in several directions.  First, we will continue
our algorithmic developments to compute the sign function of
nonhermitian matrices, with particular emphasis on novel deflation
schemes.  Second, we are currently testing the predictions of
nonhermitian random matrix theory also for the unquenched theory,
which can be done by reweighting on the small lattices that we have
studied so far.  We are also studying the average phase factor of the
fermion determinant, for which some analytical predictions exist in
the epsilon-regime of QCD \cite{Splittorff:2006fu,Splittorff:2007ck}.

\begin{acknowledgments}
  This work was supported in part by DFG grant FOR465-WE2332/4-2.  We
  would like to thank W. Bie\-ten\-holz, A. Frommer, F. Knechtli, B.
  Lang, M. L\"uscher, H. Neuberger, and J.J.M. Verbaarschot for helpful
  discussions.
\end{acknowledgments}

\bibliography{biblio}

\end{document}